\documentclass{aa}
\usepackage{graphicx}
\usepackage{txfonts}

\def\mathnew{\mathsurround=0pt}
\def\simov#1#2{\lower .5pt\vbox{\baselineskip0pt \lineskip-.5pt
       \ialign{$\mathnew#1\hfil##\hfil$\crcr#2\crcr\sim\crcr}}}
\def\simg{\mathrel{\mathpalette\simov >}}
\def\siml{\mathrel{\mathpalette\simov <}}

\def\noind{\noindent}

\def\beq{\begin{equation}}
\def\enq{\end{equation}}
\def\bea{\begin{eqnarray}}
\def\ena{\end{eqnarray}}
\def\bec{\begin{center}}
\def\enc{\end{center}}

\def\blist{\begin{list}{$\bullet$}{\itemsep 0.0in \parsep 0.0in}}
\def\elist{\end{list}}
\def\bitem{\begin{list}{\arabic{enumi}.}{\usecounter{enumi} \itemsep 0.0in \parsep 0.0in}}
\def\eitem{\end{list}}
\def\cm{\hbox{~cm}}
\def\s{\hbox{~s}}

\def\TeV{\hbox{~TeV}}
\def\GeV{\hbox{~GeV}}

\def\part{\partial}

\def\m-pl{m_{Pl}}

\def\h75{h_{75}}
\def\Omh75{\Omega h^2_{75}}
\def\Omh70{\Omega h^2_{70}}

\def\fun#1#2{\lower3.6pt\vbox{\baselineskip0pt\lineskip.9pt
  \ialign{$\mathsurround=0pt#1\hfil##\hfil$\crcr#2\crcr\sim\crcr}}}

\def\mnras{M.N.R.A.S.\,}

\def\apj{Astrophys.J.\,}
\def\apjl{Astrophys.J.Lett.\,}
\def\apjs{Astrophys.J.Supp.\,}

\def\nat{Nature\,}

\def\prd{Phys.Rev.D\,}
\def\araa{Annu.Rev.Astron.Astrophys.\,}
\def\aap{Astron.Astrophys.\,}

\def\physrep{Phys.Rep.\,}

\def\aap{\ {A\&A}\ }

\def\apj{\ {ApJ}\ }
\def\apjl{\ {ApJL}\ }
\def\apjs{\ {ApJS}\ }

\def\araa{\ {ARA\&A}\ }

\def\mnras{\ {MNRAS}\ }
\def\nat{\ {Nat}\ }

\def\physrep{\ {PhysRep}\ }

\def\prd{\ {Phys. Rev. D.}\ }

\begin{document}

\title{Gamma-Ray Bursts: Theoretical Issues and Developments}

\author{P. M\'esz\'aros }
%\inst{1}

\institute{Department of Astronomy \& Astrophysics, Department of Physics, \\
Center for Particle and Gravitational Astrophysics, Institute for Gravitation and the Cosmos,\\
Pennsylvania State University, University Park, PA 16802, USA 
\\ Email: nnp@psu.edu}

\abstract{
I discuss some aspects of the evolution of the standard GRB model, emphasizing various
theoretical developments in the last decade, and review the impact of some of the most 
recent observational discoveries and the new challenges they pose in the expanding
realm of multi-messenger astrophysics.
} 
\maketitle

\section{Genesis of the Fireball Shock Model}
\label{intro}
Fireballs in astrophysics generally refer to an optically thick plasma whose temperature 
exceeds the electron rest mass and which can produce $e^\pm$ pairs and photons in equilibrium
with a baryonic plasma.
An early study of the fireball radiation physics aimed at GRBs, leaving aside consideration 
of specific sources, was that of \cite{Cavallo+78}. The fireball would expand and adiabatically 
cool as it converts its internal into kinetic energy, and they suggested that this kinetic energy 
could be reconverted into radiation as it impacts the external medium, the highest efficiency (for 
non-relativistic expansion) being achieved when the fireball swept up an amount of external matter 
comparable to the fireball mass (the analog of the start of the Sedov-Taylor phase of SNe). 
\cite{Paczynski86} proposed that a merging binary neutron star (BNS) would liberate enough 
energy in a short time to power a GRB at cosmological distances, and he and \cite{Goodman86grb}
showed that in this case the expansion would be relativistic, the bulk Lorentz factor accelerating
with radius as $\Gamma \propto r$, The initial blackbody plasma temperature would be of order a few 
MeV, which in the linearly expanding comoving frame would drop as $T'\propto 1/r$, but in the observer 
frame this would be boosted by the bulk Lorentz factor back to its initial few MeV value, with an
approximately blackbody spectrum, most of the photons escaping  when the plasma became optically
thin to Thompson scattering. 
A different aspect of BNS  mergers emphasized by \cite{Eichler+89bns} was their 
role as emitters of gravitational waves and their likely role as sources of r-process heavy 
elements, at the same time as being likely to appear as GRBs.
A more detailed study of the properties of relativistically expanding fireballs 
\citep{Paczynski90wind,Shemi+90} 
showed that the bulk Lorentz factor growth $\Gamma(r)\propto r$ would saturate to a maximum value 
$\eta \sim E_f/M_f c^2 \gg 1$, where $E_f$ and $M_f$ are the initial energy of the explosion and the 
initial baryonic mass entrained in the outflow. After a saturation radius $r_{sat}\sim r_0 \eta$,
where $r_0$ is the launching radius, the Lorentz factor remains constant, but since adiabatic cooling 
continues, the radiation energy that can escape after the photosphere becomes optically thin represents 
an increasingly smaller fraction of the final kinetic energy of expansion. 
The dynamics is similar also for neutron star-black hole (NS-BH) mergers, which would also be 
important GRB candidates \citep{Narayan+91nsbh,Narayan+92merg}, the latter mentioning briefly that
reconnection, ejection of cosmic rays and their collisions might contribute non-thermal radiation 
in addition to the optical thick spectrum.

There were several problems with the above initial fireball models, namely, (1) for simplicity, 
a spherical geometry was usually tacitly assumed, and this combined with a low radiative efficiency 
would require excessively large explosion energies for the brighter bursts;
(2) the main part of the gamma-ray spectrum predicted is approximately blackbody, whereas observed 
spectra are mainly non-thermal; and 
(3) for plausible baryon loads most of the explosion energy would be wasted on bulk kinetic energy, 
instead of radiation.

To address these issues the jet-like fireball shock model was developed, which in its main features
is to this day the most widely used  model. As a natural way to resolve the inefficiency of 
spherical models, \cite{Meszaros+92tidal} pointed out that collimation of the fireball would be 
expected in the slower outflow (the dynamical ejecta) resulting from the tidal heating 
and the radiation from the merging BNS system. This could be powered  by reconnection between their 
magnetospheres and collisions between their winds, as well as by neutrino-antineutrino interactions 
going into pairs, which would occur preferentially along the symmetry axis of the merger. This would 
create a hot radiation bubble, which would escape through the wind preferentially along the centrifugally 
rarefied axis of rotation, making a relativistic jet.
For the case of NS-BH binary merger, \cite{Meszaros+92entropy} discussed the increased radiative 
efficiency due to gravitational focusing by the BH of the neutrino-antineutrino interactions from 
the disrupted NS debris, giving a quantitative discussion of channeling into a jet along the axis,

To address the problem of the thermal spectrum and the radiative inefficiency at the photosphere due 
to most of the energy being converted into kinetic energy form, \cite{Rees+92shock} showed that both of 
these issues are solved by considering the strong forward and reverse shocks  produced in the deceleration 
of the relativistic ejecta by the external medium, which (unlike in the non-relativistic expansion) occurs 
when the ejecta has swept up an external mass which is $\sim 1/\Gamma$ of its own mass, re-thermalizing
about half of the bulk kinetic energy. The strong shock leads to a power-law relativistic electron
spectrum via the Fermi mechanism, and via synchrotron radiation results in non-thermal power-law spectra. 
For a brief (impulsive) initial energy input, the effects of the deceleration are felt on a timescale 
$t_{dec}\sim r_{dec}/2\Gamma_i c^2$, when the initial forward shock Lorentz factor has dropped to
$\sim \Gamma_i/2$ and the reverse shock, initially weak, has just become trans-relativistic.
The results are the same whether the outflow is jet-like or spherical, for jet opening angles larger 
than $1/\Gamma$. At the photospheric radius a thermal spectrum is still emitted, but occurring above 
the saturation radius, adiabatic cooling makes its spectral contribution sub-dominant.
The dynamics and the synchrotron and inverse Compton spectra from the forward and reverse
external shock were discussed in detail in \cite{Meszaros+93gasdyn,Meszaros+93multi}.

A major motivation for introducing internal shocks arose after the launch of the Compton GRO (CGRO)
spacecraft in late 1991, which found gamma-ray light curves which showed variabilities as short as
$10^{-3}\s$. Such short variability can get smeared out in external shocks, which occur at relatively 
large radii. \cite{Rees+94is} showed that internal shocks at radii much smaller than those of the 
external shock can arise due to irregularly ejected gas shells of different bulk Lorentz factors.
These can collide and shock at intermediate radii above the photosphere but below the external shock,
leading to observable radiation whose variability is due the variability of the ejection 
from the central engine. Being above the photosphere, the shock radiation from synchrotron and
inverse Compton is unsmeared and non-thermal.

A different power source for GRBs was proposed by \cite{Woosley93col}, in addition to BNS and
NS-BH mergers.This is the collapsar model, resulting from the collapse of the core of massive 
stars leading to a central black hole (or temporarily a magnetar). When the core is rotating fast enough, 
the mass fallback towards the BH would lead to an accretion disk powering a jet, which if fed long enough, 
can break out from the collapsing stellar envelope. The BH, accretion disk and jet resulting from 
this is similar to those expected in compact BNS or NS-BH mergers, and the shock radiation outside 
the envelope would have similar properties. However the accretion can last much longer, since fall-back 
times are long and the outer accretion radii would be larger, leading to longer total burst durations.
This led to a natural explanation  for the striking dichotomy between the two populations of 
short ($\Delta t_\gamma \siml 2\s$ and long ($2\s \siml \Delta t_\gamma \siml 10^3\s$) GRBs
identified by \cite{Kouveliotou+93}.

Multi-wavelength, broadband spectra are expected in general from the external forward and reverse 
shocks, the reverse shock synchrotron predicting optical/UV radiation and the forward shock inverse 
Compton scattering of synchrotron photons reaching GeV energies \cite{Meszaros+93multi}. 
The latter provided a model \citep{Meszaros+94geV} for the long-lasting GeV emissions first seen 
in CGRO-EGRET data, while the former provided an explanation for the optical
``prompt" optical emission first detected by \cite{Akerlof+99}. Internal shocks also lead to
broadband spectra \citep{Papathanassiou+96is}, typically harder than in external shocks, due to the 
larger comoving magnetic fields at the smaller radii. 
The observation of some of the longer duration bursts, whose duration could significantly exceed
the expected deceleration time $t_{dec}\sim r_{dec}/2\Gamma_i^2 c^2$, motivated a more detailed 
discussion of external shocks \citep{Sari+95hydro,Sari97thickshell}, distinguishing between thin 
shell cases (the limiting case from brief impulsive accretion) and thick shell cases (for longer 
accretion), where the reverse shock may  be relativistic.

The long-term afterglow of a GRB, as opposed to the ``prompt" emission discussed above, was first 
discussed quantitatively \citep{Meszaros+97ag} in a paper which appeared two weeks before the first
announced detection of an X-ray afterglow from GRB 970228 with the Beppo-SAX satellite 
\citep{Costa+97-970228}. Its optical afterglow was discovered by \cite{Vanparadijs+97-gtrb970228opt}, 
and other afterglows soon followed, including GRB 970508 \citep{Metzger+97-970508redshift} which 
yielded the first redshift ($z=0.835$), proving that they were indeed cosmological.
The observations confirmed in their main features the predictions of the afterglow model, including
the power law time decay, spectra  and timescale \citep{Wijers+97-970228}.  Synchrotron radiation, 
including the transition between slow and fast cooling regimes \citep{Meszaros+98view,Sari+98spectra},  
provided a satisfactory fit for most of the observations in the subsequent period.

\section{Standard GRB Model and its Evolution}
\label{sec:sm}
\begin{figure}[h]
   \centering
   \includegraphics[width=4cm]{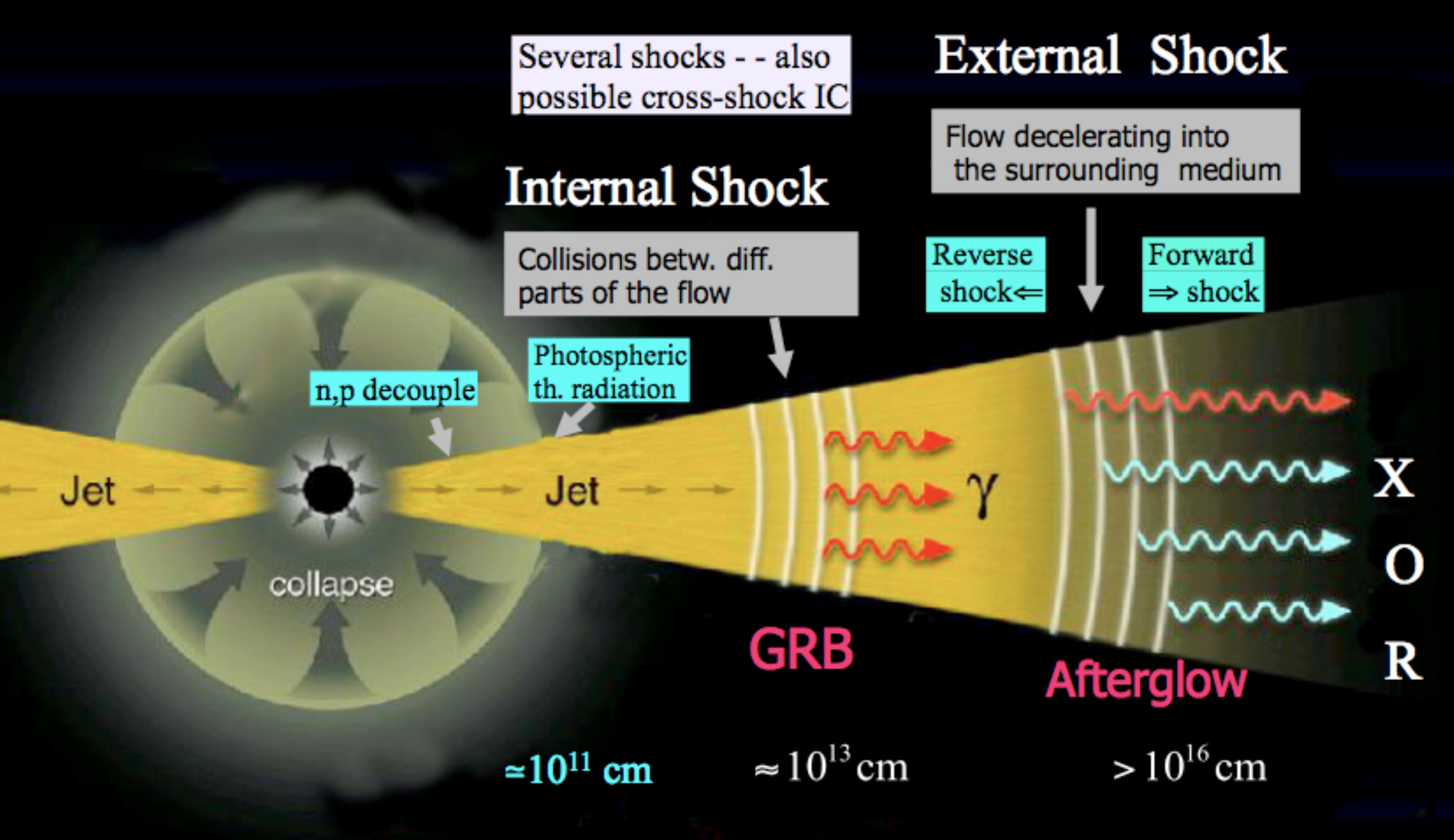}
   \caption{The standard GRB fireball shock model, e.g. from a collapsar
            (for compact mergers, the ``collapse" region is replaced by
             the dynamical ejecta). Shown are the photosphere, internal shock 
             and external shock resulting in the afterglow \citep{Meszaros01sci}.
               }
              \label{fig:smjet}%
    \end{figure}

The ``standard" GRB fireball shock model outlined in the second half of the above section
has proved extremely durable, despite a number of challenges and modifications of detail.
For comprehensive reviews, see e.g., \cite{Piran99rev,Piran04grbrev,Kumar+15grbrev,Zhang19book}.
Its simplest form, most often used for interpreting observations, is in Fig. \ref{fig:smjet}.

The afterglow radiation, from radio through optical, X-ray and more recently GeV is overall well 
fitted, with some modifications, by the external shock synchrotron emission, e.g. \cite{Zhang+06ag}.
For the ''prompt" emission (broadly the typically MeV radiation within $\Delta t_\gamma \simeq T90$), 
however, an origin in terms of synchrotron has been criticized, e.g. \cite{Preece+98death}, since 
the low energy slope of some GRB prompt spectra is harder then the limiting synchrotron slope of 
-2/3 in $dN/dE$ (harder than +1/3  in $E dN/dE$). 

\begin{figure}[h]
%\vspace*{-0.15in}
%\begin{minipage}[h]{0.5\textwidth}
\begin{minipage}[h]{3.5cm}
\centerline{\includegraphics[width=3.5cm,height=3.0cm]{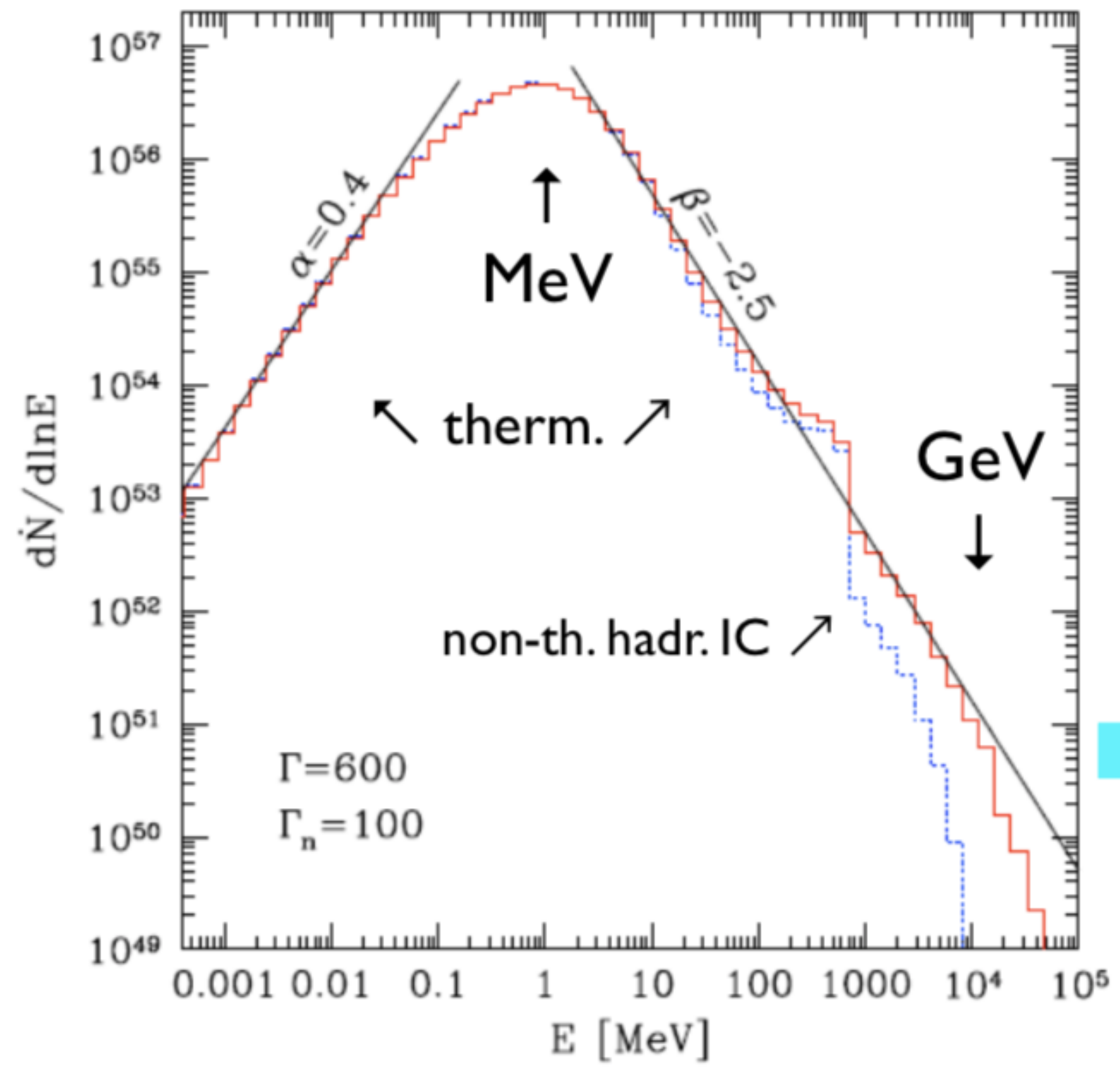}}
\end{minipage}
%\hfill
%\hspace{15mm}
\vspace*{-3cm}
%\begin{minipage}[h]{0.5\textwidth}
\begin{minipage}[h]{3.5cm}
\centerline{\includegraphics[width=3.5cm,height=3.0cm]{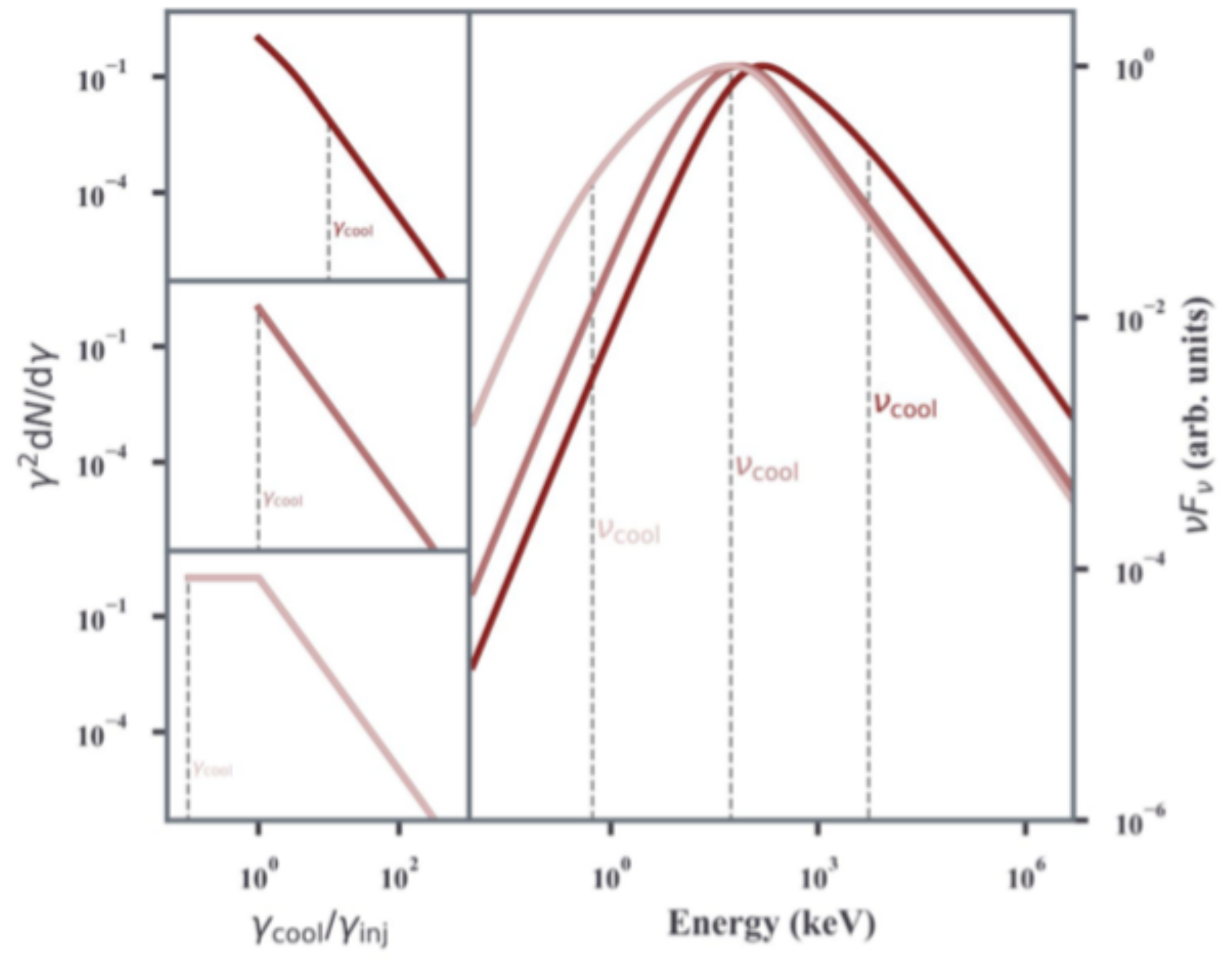}}
\end{minipage}
\vspace*{3cm}
\caption{Left: Spectrum of a photosphere heated by $pn$ decoupling \citep{Beloborodov10pn}.
Right: Synchrotron spectra (e.g. internal or external shocks) accounting for  time-dependence of 
transition between cooling regimes \citep{Burgess+18grbsync}.  }
\label{fig:phot-sync}
\end{figure}

One possible solution is that the prompt emission
may be due to the optically thick photosphere, whose peak can be in the MeV range and the low
energy slope is as hard as +2 \citep{Eichler+00thermal}. This works but it requires an additional 
shock or other component to make a high energy power law  \citep{Meszaros+00phot}; also,
if the photosphere is well above the saturation radius adiabatic cooling makes it radiatively
inefficient. Radiatively efficient photospheres, however, may arise naturally if the photospheres 
are dissipative \citep{Rees+05photdis}, e.g. by magnetic reconnection, or subphotospheric shocks.
A natural sub-photospheric dissipation mechanism is proton-neutron decoupling, which can produce
efficient photospheric spectra from low energies all the way to multi-GeV \citep{Beloborodov10pn},
Fig. \ref{fig:phot-sync} (left).

However, the earlier critiques of the synchrotron low energy slopes may be unjustified, having been
obtained taking wide time bins or time-integrated spectra. If one considers the time evolution
of the shock-accelerated electrons, from injection through cooling, and takes into account 
that electrons radiating at different frequencies have different energies and may be in
different cooling regimes (fast, intermediate, slow), the spectra convolved with the detector
energy resolution and response function can give various slopes, and the great majority of
the observed GRB slopes can be fitted with synchrotron, e.g. \cite{Burgess+18grbsync,
Ravasio+19sycoolspec}, Fig. \ref{fig:phot-sync}, right.

\begin{figure}[h]
   \centering
   \includegraphics[width=4cm]{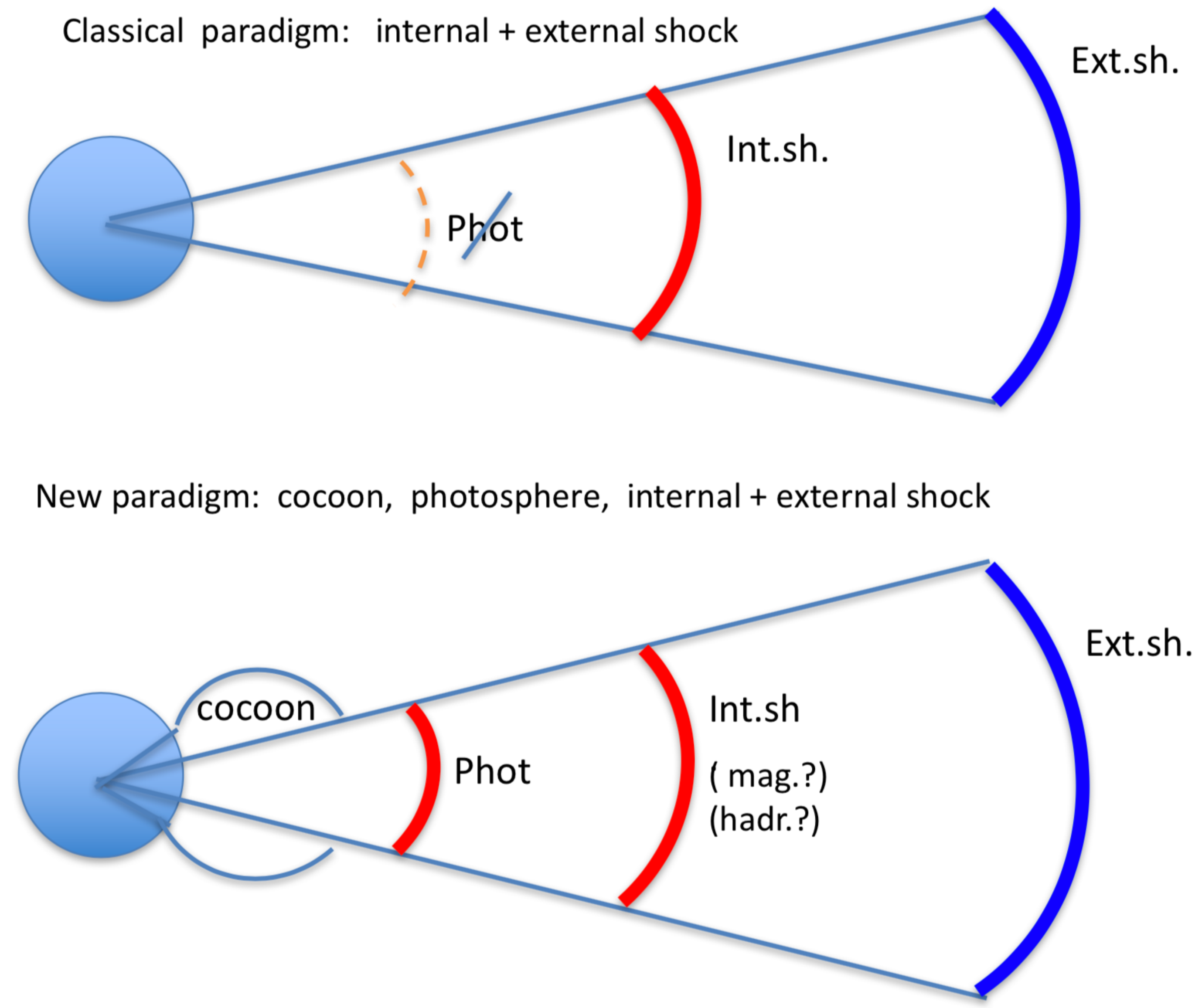}
   \caption{The early classical paradigm of the standard model (top) and
            the newer version giving more emphasis to the photosphere
            and considering alternative mechanisms in the internal shock
            or prompt emission region.
               }
              \label{fig:grb-paradigm}%
    \end{figure}

A critique of the simple internal shocks in which only the electrons radiate (leptonic models)
has been that they are generally inefficient, not dissipating enough of the mechanical energy 
in the relative motion between successively ejected shells, and Fermi acceleration putting 
much of this dissipated energy in non-radiating protons. In more realistic internal shocks, however, 
this radiative inefficiency can be larger, e.g. when the dissipation is largely by magnetic 
instabilities and reconnection, or if hadronic collisions and reacceleration of secondaries are taken 
into account. Thus, the early GRB paradigm based on internal plus external shocks and an inefficient
photosphere (Fig. \ref{fig:grb-paradigm}, top) has, since about 2005, evolved into one of an 
efficient photosphere and/or an efficient internal shock plus external shock 
(Fig. \ref{fig:grb-paradigm}, bottom).
An example of efficient magnetic dissipation internal shock models is the ICMART model 
of \cite{Zhang+11icmart}, while an example of an efficient hadronic internal shock with 
secondary reacceleration is that of \cite{Murase+12reac}. 

After the launch of the Fermi Gamma-ray Space Telescope 2008, its LAT detector started
to observe in a significant fraction of GRBs that the so-called Band %\citep{Band+93}
broken power law spectrum above the MeV peak extended into the GeV range, as already found 
in some previous EGRET spectra. Such ``extended"  Band spectra can be modeled, e.g. with 
photospheric models, as seen in Fig. \ref{fig:phot-sync} (left). However, in many Fermi-LAT 
GRBs the GeV appeared as a second power-law component, harder than the Band $\beta$ upper branch. 

\begin{figure}[h]
%\vspace*{-0.15in}
%\begin{minipage}[h]{0.5\textwidth}
\begin{minipage}[h]{3.5cm}
\centerline{\includegraphics[width=3.5cm,height=3.0cm]{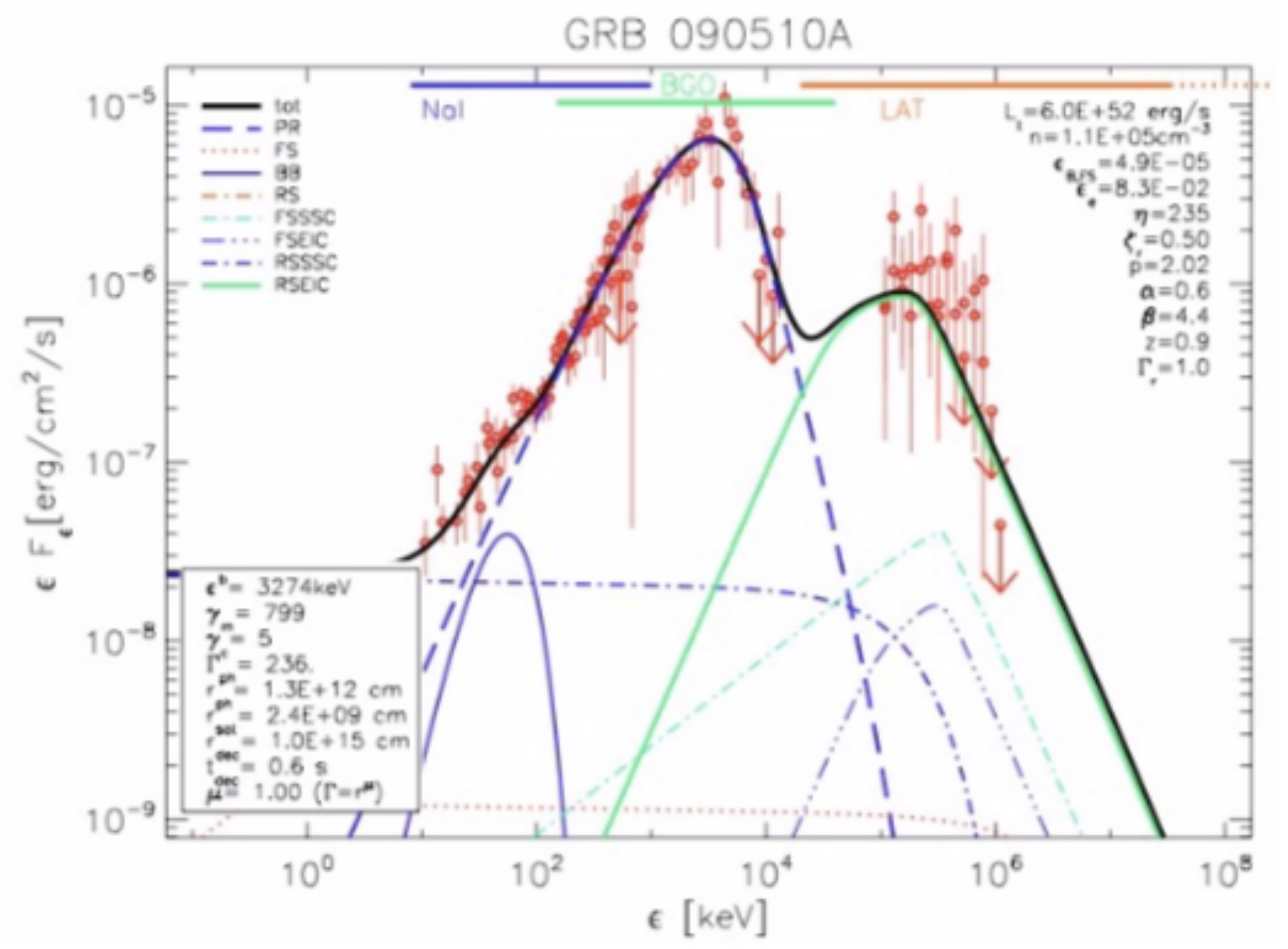}}
\end{minipage}
%\hfill
%\hspace{15mm}
\vspace*{-3cm}
%\begin{minipage}[h]{0.5\textwidth}
\begin{minipage}[h]{3.5cm}
\centerline{\includegraphics[width=3.5cm,height=3.0cm]{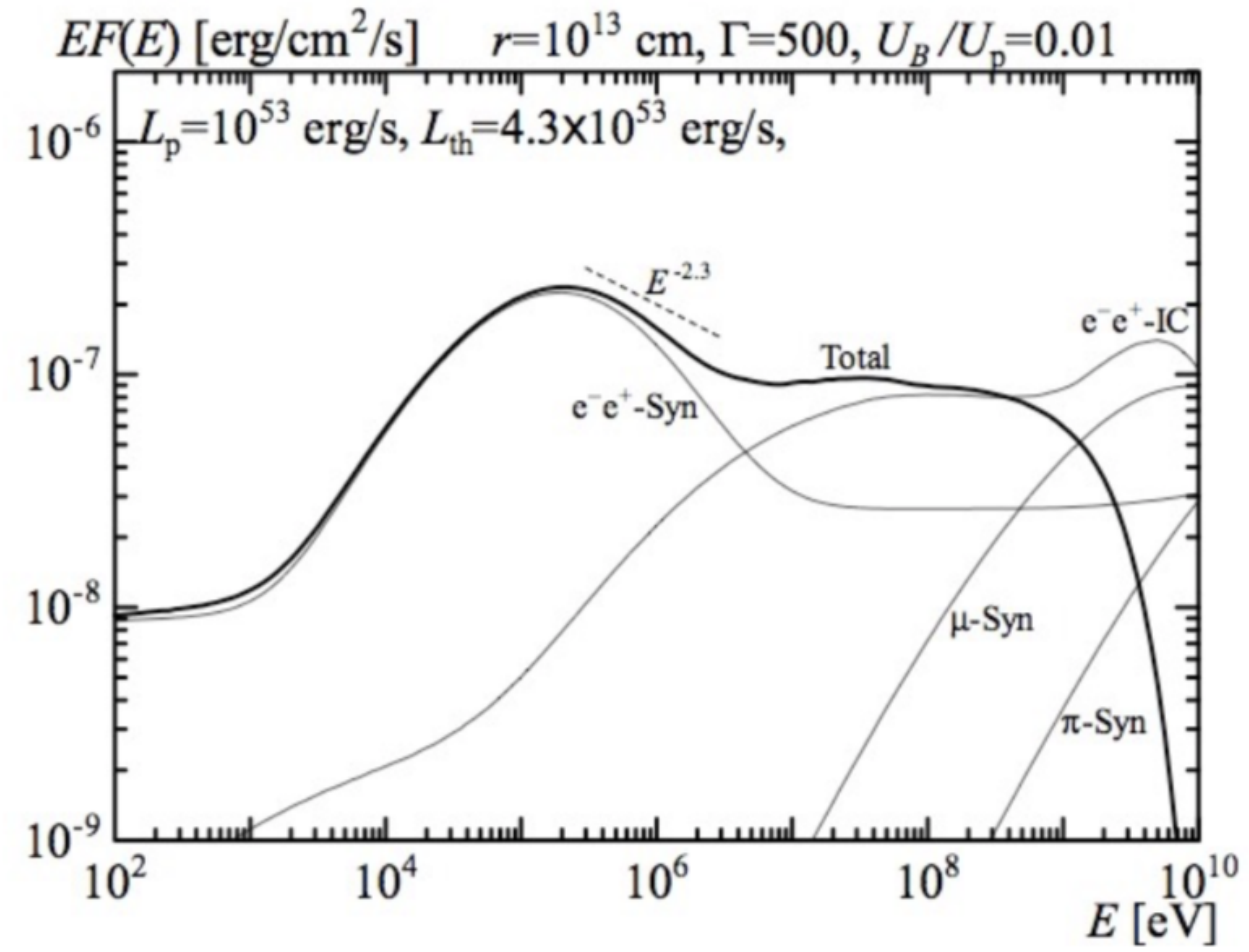}}
\end{minipage}
\vspace*{3cm}
\caption{Left: Leptonic model with photosphere plus external shock upscattering a GeV second 
component \cite{Veres+12photes}.
Right: Hadronic model with internal shock accelerating electrons and protons leading to cascades 
and secondary re-acceleration, leading to self-consistent Band spectrum and second GeV component
\citep{Murase+12reac}.}
\label{fig:gevlephad}
\end{figure}

The question is whether this second, harder GeV component is
due to inverse Compton (IC) upscattering of the Band component, or is it due to protons being 
accelerated and leading to cascades with radiation from secondary leptons. Both types
of models can give reasonable fits. Leptonic models where the Band spectrum arising in the 
photosphere is up-scattered by shocked electrons in internal shocks give reasonable results, 
e.g.  \cite{Toma+11phot}. An  alternative leptonic model considers a baryonic or magnetic 
photosphere producing a Band spectrum which is up-scattered in an external shock, also giving
good fits \citep{Veres+12photes}. Fig. \ref{fig:gevlephad} (left).
\\
Hadronic models, on the other hand, could in principle have substantial advantages.  E.g. 
\cite{Murase+12reac} calculated an internal shock model accelerating both electrons and protons,
where hadronic cascades and stochastic reacceleration of the leptonic secondaries in the
post-shock turbulence leads self-consistently to both the Band MeV and the GeV second hard component 
from the same region, Fig. \ref{fig:gevlephad} (right). This model provides good efficiency, which 
is one of its attractions for internal shocks, and it may be applicable not only to shocks but also, 
e.g. to magnetic dissipation regions, where MHD turbulence is expected.

\section{Some Recent Developments}
\label{sec:recent}
\begin{figure}[h]
%\vspace*{-0.15in}
%\begin{minipage}[h]{0.5\textwidth}
\begin{minipage}[h]{3.5cm}
\centerline{\includegraphics[width=3.5cm,height=3.0cm]{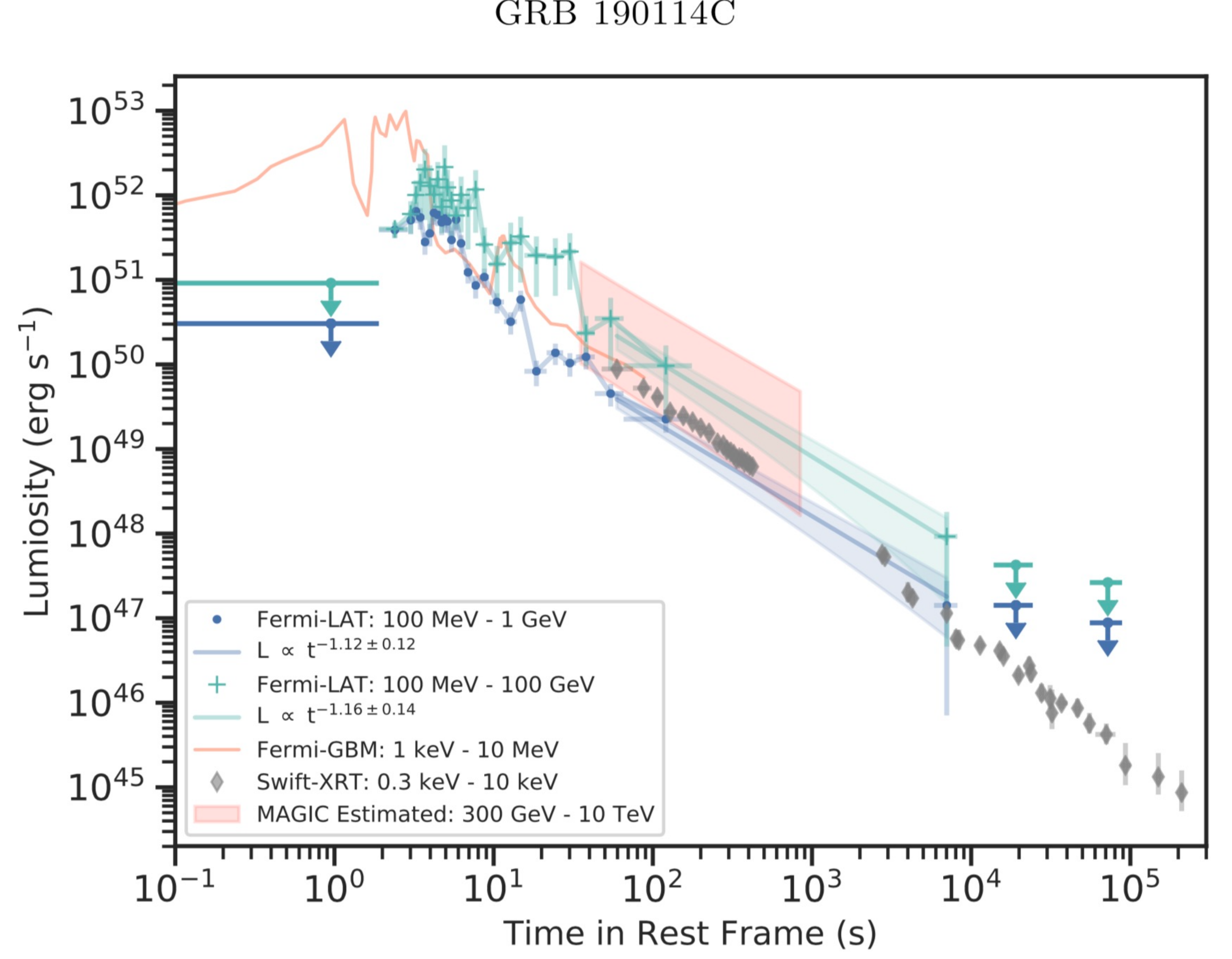}}
\end{minipage}
%\hfill
%\hspace{15mm}
\vspace*{-3cm}
%\begin{minipage}[h]{0.5\textwidth}
\begin{minipage}[h]{3.5cm}
\centerline{\includegraphics[width=3.5cm,height=3.0cm]{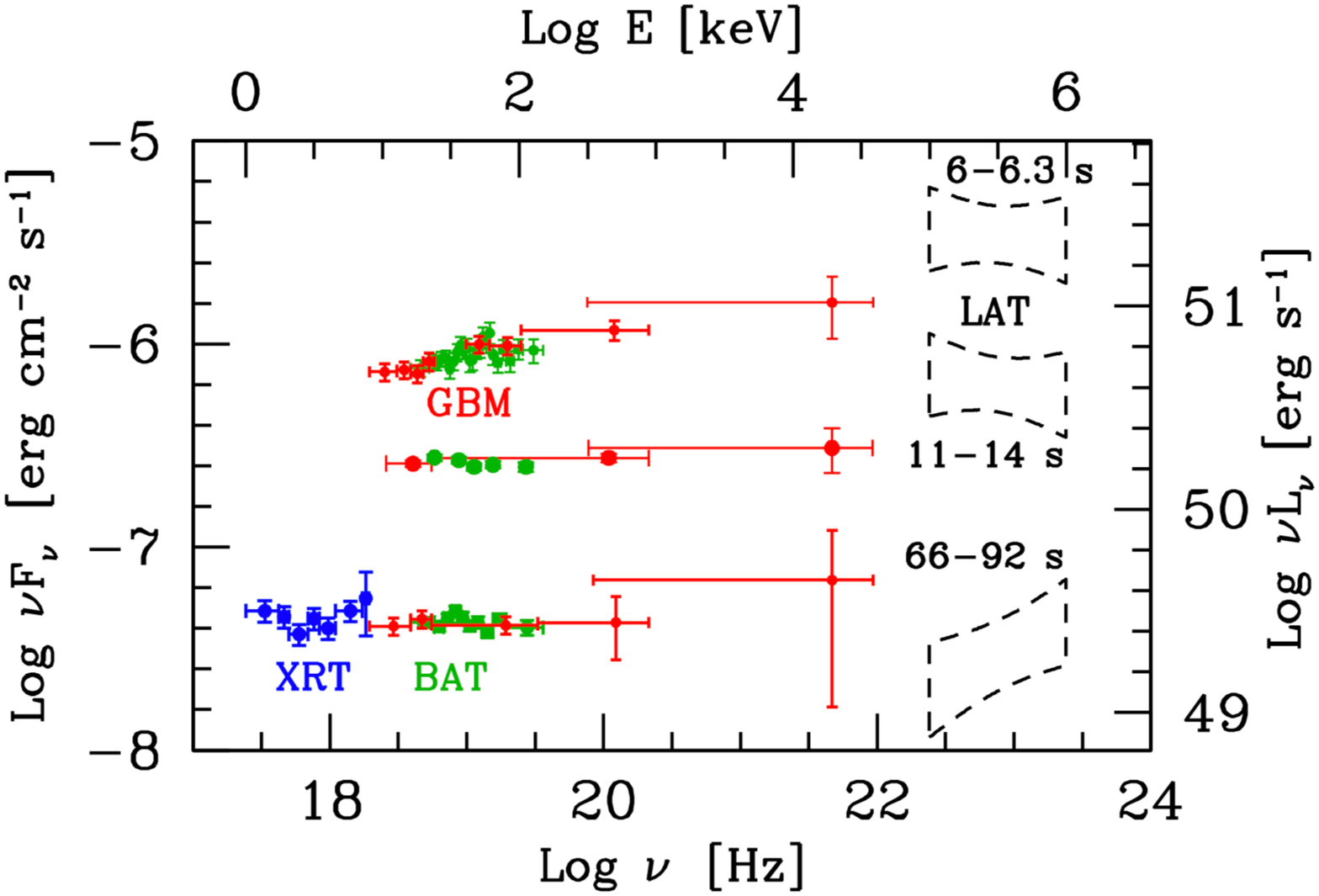}}
\end{minipage}
\vspace*{3cm}
\caption{Left: fits to preliminary data showing light curves of various energy components of GRB 190114C
\citep{Wang+19-190114tev}. 
Right: fits to preliminary data at several epochs for the spectrum of GRB190114C 
\citep{Ravasio+19-190114Cspec}.}
\label{fig:190114Clcspec}
\end{figure}

Recently the MAGIC imaging air fluorescence telescope (IACT) announced the detection of photons in 
excess of 300 GeV, and perhaps up to a TeV, in the bright GRB190114C \citep{Mirzoyan+19-190114C-MAGIC}, 
also detected at other energies by Fermi, Swift, INTEGRAL and numerous other facilities. This was the
first high confidence ($\sim 20\sigma$) detection of a GRB  with an IACT at such energies, a long 
awaited feat which should be easier to accomplish with the future CTA. Preliminary analyses show that 
the long lasting  ($\sim 10^3\s$) sub-TeV component is mostly associated with the afterglow seen at 
other energies, e.g. Fig. \ref{fig:190114Clcspec}. The spectral slope of the sub-TeV component 
appears  harder than the usual Band component, pending further MAGIC analysis. 
\\
Detection of $\siml \TeV$ emission from a GRB  has two requirements, one being that the redshift 
be smallish, so that $\gamma\gamma$ absorption in the IGM external background light is not too
severe, and the other being that the same absorption is absent or at least mitigated in the GRB radiation 
zone and its immediate neighborhood. Fermi-LAT detections of GRBs have shown source-frame emission 
up to several tens of GeV and in one case even $\siml 100\GeV$, but the present $\simg 300\GeV$ 
can put significant constraints on models. Much theoretical work remains to be done on this  event.

The other major recent development was the short GRB 170817 detection both electromagnetically (EM) through 
multi-wavelength photons, and through gravitational waves (GWs). This was very exciting, being the first 
high significance multi-messenger detection of a transient using GWs\footnote{The other previous high 
significance multi-messenger transient was SN 1987a, where besides photons also thermal (MeV) neutrinos 
were detected.}.
%\cite{fig:170817-lc-schem}
%
This was a short GRB (SGRB) with $\Delta t_\gamma \leq 2\s$), detected by Fermi, INTEGRAL, Swift 
and other EM instruments. These objects were long expected to arise from BNS mergers, an interpretation 
for which accumulating evidence, e.g. \cite{Gehrels+09araa}, had almost but not quite reached the 100\% 
confidence level. In this case, slightly preceding the EM flash, the associated detection of GWs 
\citep{LIGO+17gw170817disc}, which were also expected from BNS mergers, conclusively confirmed that 
SGRBs where indeed BNS mergers. In addition, it also confirmed that BNSs can also produce a type of 
optical/IR flash known as a kilonova, surmised to be responsible also for the elements heavier than the 
Fe-group via the r-process, e.g.  \cite{Hotokezaka+18nsmerg,Kasliwal+19-170817rproc}..
\begin{figure}[h]
\centering
\includegraphics[width=7.0cm,height=3.5cm]{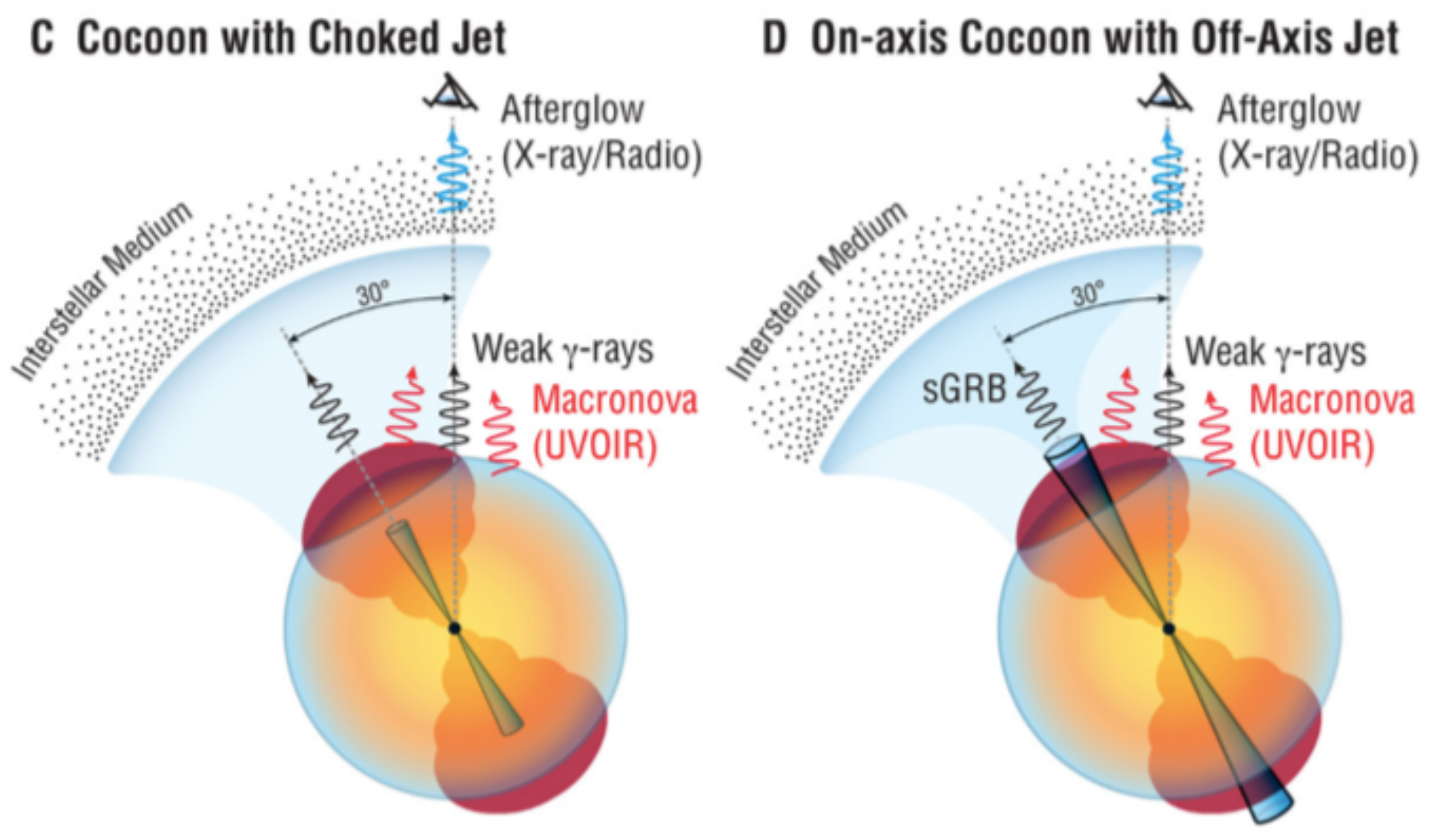}
\caption{GRB/GW170817A, two opposed views on the SGRB radiation from a BNS.
Left: observed radiation dominated by the cocoon, for a choked jet. 
Right: observed radiation dominated by an emergent top-hat jet \citep{Kasliwal+17-170817sci}.}
\label{fig:Kasliwal17cocjet}%
\end{figure}

The SGRB radiation of GRB/GW170817 looked typical, except for being fainter and somewhat softer than 
expected for its low distance of 40 Mpc. The role played in GRBs by cocoons \citep{Meszaros+01coljet, 
Ramirez+02cocoon} and choked jets \citep{Meszaros+01choked} had been considered early on, and in 
the case of GRB/GW170817A a natural possibility was that its weaker $\gamma$-rays might be attributed 
either to a choked jet with a cocoon breakout, or to an off-axis top-hat emergent jet, e.g.
\cite{Kasliwal+17-170817sci,Ioka+17gw170817cocoon} and others. While a cocoon interpretation may
be favored over a simple top-hat jet, the observation of a superluminal jet signature 
\citep{Mooley+18-170817reljet2} and other features of the afterglow  \citep{Troja+18-179817structjet}
indicate that either a Gaussian structured jet or a cocoon could fit the data.

The next burning question, as far as GRB multi-messenger studies, is whether GRBs can also be
detectable via neutrinos. Of course both LGRBs (as core-collapse objects) and SGRBs (compact 
mergers involving at least one neutron star which is heated to virial temperatures) will emit a 
large fraction of their core binding energy in thermal (5-30 MeV) neutrinos. At these energies the 
neutrino-nucleon detection cross section is of order $10^{-44}\cm^2$, and at cosmological distances 
the flux is undetectable with current detectors. High energy neutrinos however have much higher cross 
sections ($\sim 10^{-34}\cm^2$ around 10 TeV), and IceCube is detecting a diffuse astrophysical flux 
in the 10 TeV-10 PeV range \citep{IC3+13pevnu1,IC3+13pevnu2}. The total number of neutrinos so far 
is of order 50, distributed isotropically in the sky, with localization error circles ranging from
$\sim 1^o$ (for muon neutrino tracks) to $15-30^o$ (for electron neutrino cascades), hence difficult 
to associate with individual sources. Recently, however, a high energy (multi-TeV) muon neutrino 
was detected, with the blazar TXS 0506+56 within its error circle, which was undergoing a $\gamma$-ray
flaring episode in near time coincidence with the neutrino arrival. The region also showed other 
previous neutrinos in the past years, but without coincident $\gamma$-ray flares, so the total
coincidence significance is $\sim 3.5\sigma$, which is interesting but not yet considered 
conclusive evidence \citep{IC3multi+18txs0506,IC3+18txs0506previousnu}.

The possibility of GRBs being high energy neutrino sources has been investigated by IceCube 
using classical GRBs, i.e. bright, EM-detected, mainly LGRBs. These have been 
disfavored by IceCube analyses, e.g. \cite{IC3+15promptnugrb}, using particular models of the neutrino 
emission expected.  The same conclusion is reached by IceCube for classical GRBs in a model-independent 
way using constraints based on neutrino multiplet observations \citep{IC3+18grbnuconstraint}, but the 
same study leaves unconstrained a (theoretically plausible) origin in low-luminosity or choked GRBs, 
e.g.  \cite{Senno+16hidden}. Low luminosity and/or choked GRBs could be more numerous than classical 
GRBs, and at the typically high redshifts they would be electromagnetically missed or hard to detect, 
while their cumulative neutrino flux could add up to what IceCube sees.

\begin{figure}[h]
%\vspace*{-0.15in}
%\begin{minipage}[h]{0.5\textwidth}
\begin{minipage}[h]{3.5cm}
\centerline{\includegraphics[width=3.5cm,height=2.5cm]{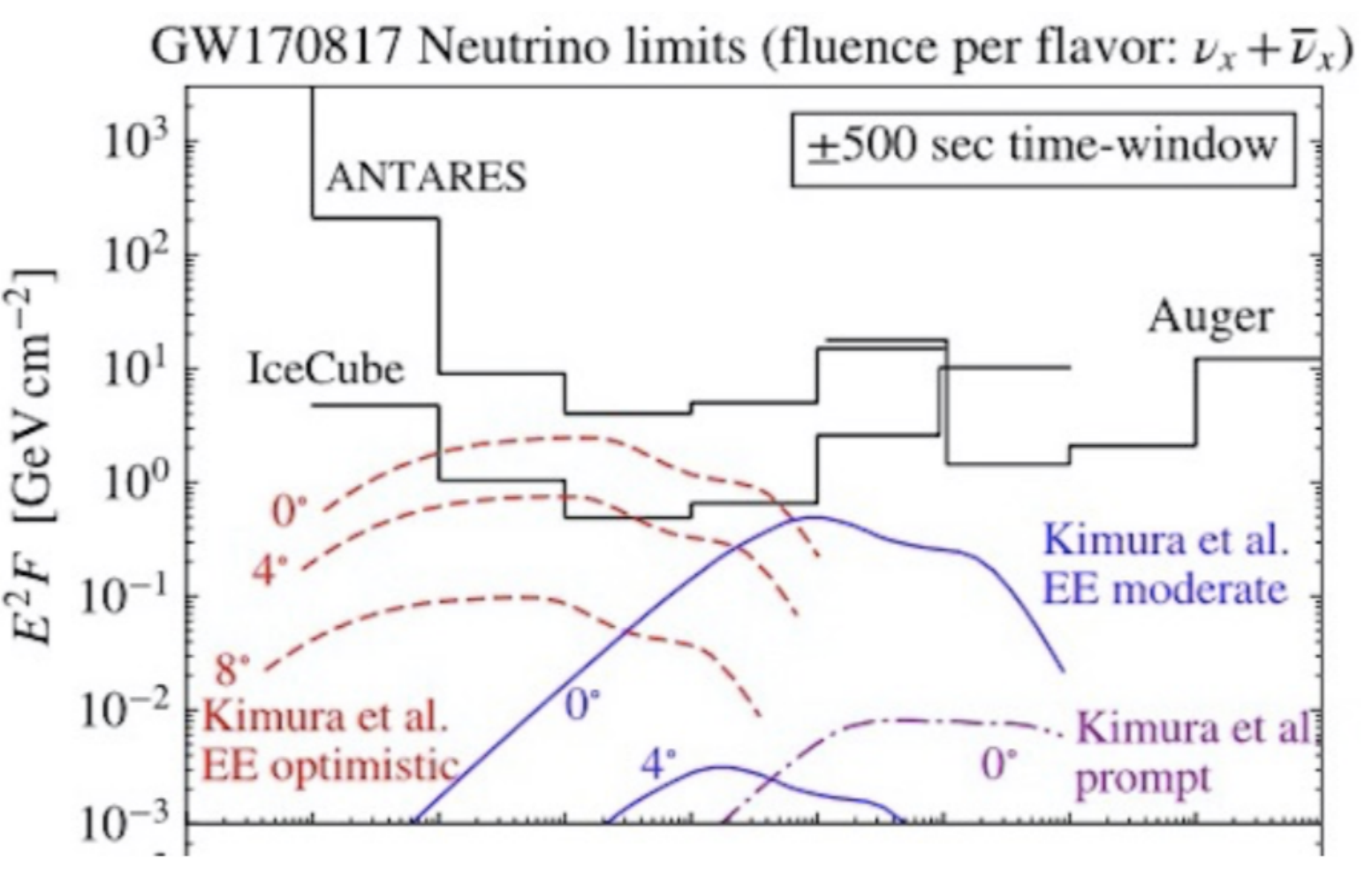}}
\end{minipage}
%\hfill
\hspace{5mm}
\vspace*{-3cm}
%\begin{minipage}[h]{0.5\textwidth}
\begin{minipage}[h]{3.5cm}
\centerline{\includegraphics[width=3.7cm,height=2.5cm]{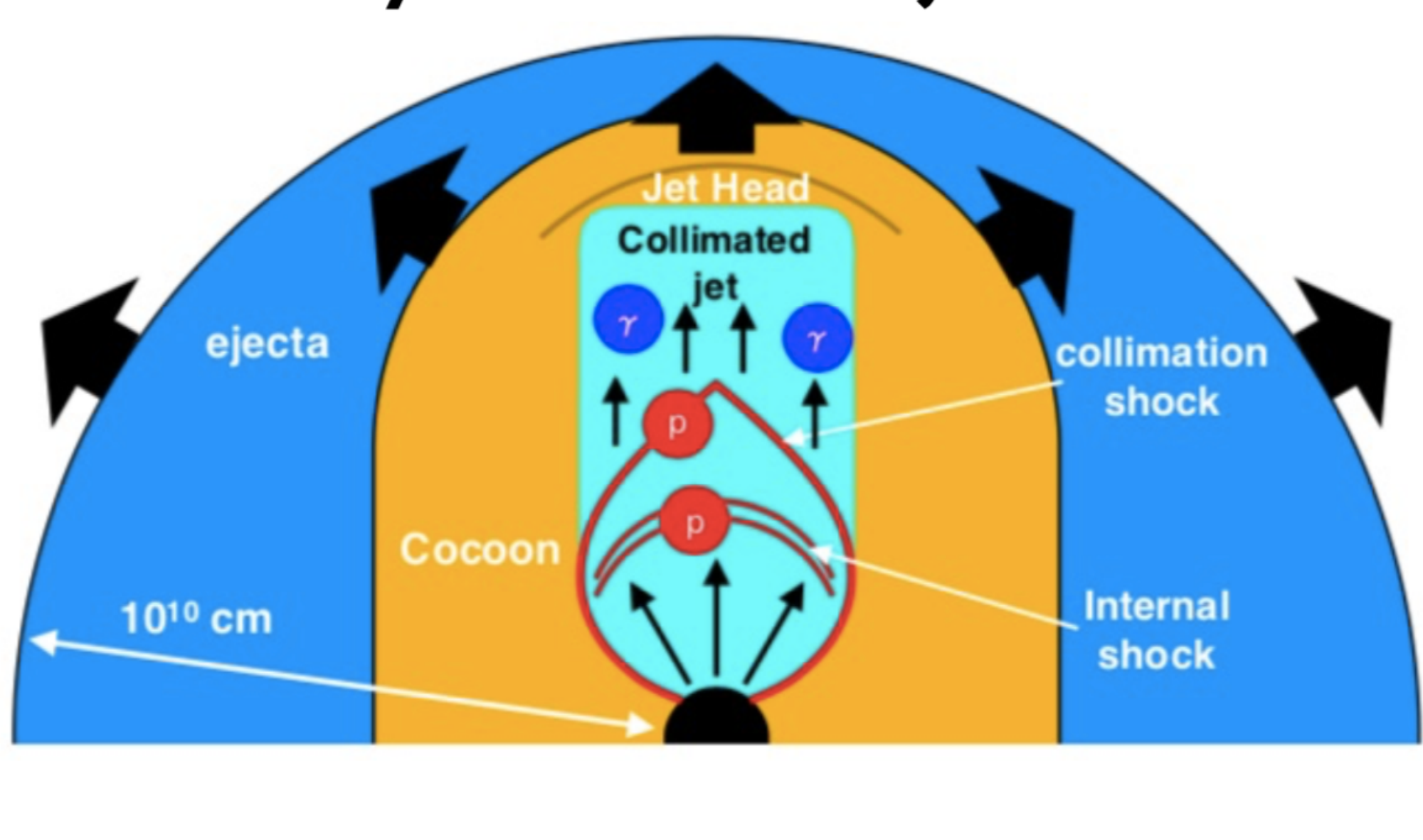}}
\end{minipage}
\vspace*{3cm}
\caption{ Left: IceCube and Antares upper limits for GRB170817 \citep{AntaresIC3-170817nu}, 
compared to BNS jet EE extended emission model \citep{Kimura+17sgrbnu} for various jet offset angles.
Right: Internal and collimation shocks in trans-ejecta jet propagating through a BNS dynamical ejecta
\citep{Kimura+18transejnu}.}
\label{fig:bnsnu}
\end{figure}

SGRBs would in principle appear to be ideal objects for constraining physical models if in addition 
to GWs they also produced observable neutrinos. At first sight it would seem that the expected
neutrino fluxes would be much lower than in LGRBs, since the SGRB prompt MeV emission is shorter and
underluminous compared to LGRBs. However a large fraction of SGRBs also exhibit a longer tail 
($\siml  100\s$) of softer radiation in the 50 keV range. This softer extended emission (EE) can be
modeled as a late jet emission with a bulk Lorentz factor lower than the prompt, providing a higher 
comoving density of target photons for $p\gamma$ photo-hadronic interactions leading to neutrinos. 
The neutrino flux is still low at typical redshifts, but at the redshift $z\sim 0.01$ (about 40 Mpc) 
of GBB 170817A, it could have been detectable by IceCube, if the jet been head-on \citep{Kimura+17sgrbnu};
however, for a higher inclination angle $\theta_{LOS} \sim 20-30^o$ of the line of sight relative 
to the jet axis, as inferred from multi-wavelength observations, the lower Doppler boost in that 
direction implies a much lower observable flux, which falls below the IceCube sensitivity, 
Fig. \ref{fig:bnsnu} (left).
\\
The SGRB jet and shock structure is likely to be more complicated as it is making its way through
the dynamical ejecta, Fig. \ref{fig:bnsnu} (right). Both collimation shocks and internal shocks are
expected in choked jets or before the jet emerges from the ejecta, and the internal shocks occurring 
in the pre-collimation jet satisfy the conditions for Fermi acceleration of charged particles, leading 
to neutrinos via photo-hadronic interactions \citep{Kimura+18transejnu}. One can expect from such
events a few up-going neutrinos in IceCube from a merger at 40 Mpc occurring in the Northern sky, 
if the jet is directed at Earth.  For optimistic jet parameters, a joint GW-IceCube detection might 
be achievable in a few years of operation, or for Ice-Cube Gen 2 this would be probable even for 
moderate jet parameters.
\\
\noind
{\it Acknowledgments:} I am grateful to K. Murase, S.S. Kimura and D.B. Fox for discussions, and the 
Eberly Foundation for support.

\bibliographystyle{aa.bst}

%\footnotesize

%\bibliography{/Users/nnp/ms/grb.bib,/Users/nnp/ms/crnu.bib,/Users/nnp/ms/gw.bib,/Users/nnp/ms/cluster.bib,/Users/nnp/ms/agnvhe.bib,/Users/nnp/ms/star.bib,/Users/nnp/ms/tde.bib,/Users/nnp/ms/mhd.bib,/Users/nnp/ms/frb.bib,/Users/nnp/ms/mag.bib,/Users/nnp/ms/phys.bib,/Users/nnp/ms/cosm.bib,/Users/nnp/ms/instr.bib,/Users/nnp/ms/cluster.bib,/Users/nnp/ms/bns.bib}

\end{document}